\documentclass[12pt,amsmath,amssymb]{iopart}
\usepackage{iopams} \usepackage{graphicx}

\begin{document}
\title{Neutrino Mixing and Nucleosynthesis in Core-Collapse Supernovae }
\author{A.B. Balantekin\dag \ddag \footnote[3]{baha@physics.wisc.edu   }
       and H. Y\"{u}ksel\dag \footnote[5]{yuksel@physics.wisc.edu}	}
\address{\dag\ Department of Physics, University of Wisconsin\\
               1150 University Avenue, Madison, WI 53706 USA\\ 		}
\address{\ddag\ Physics Department, Tohoku University  \\
Sendai 980 Japan}               

\begin{abstract}
A simple description of core-collapse supernovae is given. Properties
of the neutrino-driven wind, neutrino fluxes and luminosities, reaction
rates, and the equilibrium electron fraction in supernova environments
are discussed. Neutrino mixing and neutrino interactions that are
relevant to core-collapse supernovae are briefly reviewed. The values
of electron fraction under several evolution scenarios that may impact
r-process nucleosynthesis are calculated. 
\end{abstract}

\submitto{\NJP}
\maketitle


\section{Introduction}\label{sec:introduction}

Massive stars, losing energy to radiation and photons, evolve until an
iron core is formed (for a recent review see Ref. \cite{woos1}). This
core, which has a very low entropy per baryon, is supported by the
electron degeneracy pressure. As a consequence such a core is
dynamically unstable and collapses until the matter is mostly
neutronized and supernuclear densities are reached. Only when the
density significantly exceeds the nuclear density, the pressure
becomes sufficiently repulsive to stop the collapse.
In the current
paradigm the innermost shell of matter reaches to these densities first,
rebounds and sends a pressure wave through the rest of the
core. Such waves, produced by the subsequent shells and traveling
faster than the infalling matter, collect near the sonic point. As
that point reaches nuclear density a shock wave breaks out. The
subsequent evolution of this bounce shock is not yet well-understood
and is subject to much study \cite{Buras:2003sn,Mezzacappa:2004ic}.
Current models fail to explode.

Even though it has not yet been demonstrated that explosion is an
outcome of the core-collapse, it is well-established that the
newly-formed hot proto-neutron star cools by neutrino emission. (It
was those neutrinos that were observed in Supernova 1987A). Almost all
($99\%$) of the gravitational binding energy of the neutron star
\begin{equation}\label{eq:1}
\frac{3}{5} \frac{G\,M_{NS}^2}{R_{NS}}
\approx 3 \times 10^{53} {\rm ergs} \>
\frac{(M_{NS}/1.4M_\odot)^2}{R_{NS}/10 {\rm km}}
\end{equation}
is radiated away in neutrinos of all flavors. Altogether a star with
mass $\sim 8 M_{\odot}$ will emit $\sim 10^{59}$ neutrinos.
Thus, after the explosion ejects the material from the outer layers,
a ``neutrino-driven'' wind may blow the medium above the neutron star,
heating it to an entropy per baryon of several hundreds in units of
Boltzmann's constant. For such high entropies nuclear statistical
equilibrium is not established for nuclei heavier than alpha
particles.

The rapid neutron-capture process (r-process) is responsible for the
formation of a number of nuclei heavier than iron. (For a recent
review see Ref. \cite{Qian:2003wd}).
The astrophysical site of the r-process nucleosynthesis is not yet
identified. For r-process nucleosynthesis to successfully take
place a large number of the
neutrons are required to interact in a relatively short time,
indicating that r-process sites are associated with explosive
phenomena. Indeed the seminal Burbidge, Burbidge, Fowler, and Hoyle
paper 
suggested the neutron-rich ejecta outside the core in a type II
supernova as a possible site of the r-process \cite{b2fh}.
More recent work pointed to the neutrino-driven wind in the supernovae
as a possible site \cite{Woosley:1994ux,witti,Wanajo:2001pu}. Meteoric
data and observations of metal-poor stars indicate that r-process
nuclei may be coming from diverse sources \cite{Qian:1997kz}. Binary
neutron star systems were also proposed as a site of the r-process
(see e.g. Ref \cite{Rosswog:1998hy}). In outflow models r-process
nucleosynthesis results from the freeze-out from nuclear statistical
equilibrium. The outcome of the freeze-out process is determined by
the neutron-to-seed ratio. This ratio in the
post-core-bounce supernova environment is controlled by the intense
neutrino flux radiating from the neutron star.

Neutrino interactions play a crucial role in core-collapse
supernovae. (For a brief summary see Ref. \cite{Balantekin:2003ip}).
Neutrino heating is one of the possible mechanisms for reheating
the stalled shock \cite{Bethe:1984ux}.
The neutrino fluxes control the
proton-to-neutron ratio in the high-entropy hot bubble.
As we describe in the next
section there is
a hierarchy of energies for different neutrino flavors.
Hence swapping active neutrinos via neutrino oscillations changes
the $n/p$ ratio and may alter
r-process nucleosynthesis conditions
\cite{Qian:1993dg}. Neutrino oscillations in a core-collapse
supernova differ from the matter-enhanced neutrino oscillations in
the Sun as in the former there are additional effects coming from both
neutrino-neutrino scattering \cite{Qian:1994wh,Pantaleone:1994ns} and
antineutrino flavor transformations \cite{Qian:1995ua}.

We present a simple description of the core-collapse supernovae in the 
next Section. In this section 
after summarizing properties of the neutrino-driven wind, 
we discuss neutrino fluxes, luminosities, reaction rates, and the
equilibrium  
electron fraction. A brief description of neutrino mixing and neutrino 
interactions relevant to core-collapse supernovae is given in Section 3. 
In Section 4 we calculate values of the electron fraction under several 
evolution scenarios that may impact r-process nucleosynthesis. 

\section{A Simple Description of Core Collapse Supernovae}

\subsection{Neutrino-Driven Wind}

A careful treatment of the neutrino-driven wind in post-core bounce
supernova environment 
was given in Ref. \cite{Qian:1996xt}. Here we present a
heuristic description following Ref. \cite{McLaughlin:1999pd}.
One can assume that at sufficiently large radius above the heating
regime there is hydrostatic equilibrium \cite{Bethe:1992fq}:
\begin{equation}
\label{a1}
\frac{dP}{dr} = - \frac{G M_{\rm NS} \rho}{r^2} ,
\end{equation}
where $P$ is the hydrostatic pressure, $G$ is Newton`s constant, $
M_{\rm NS}$ is the mass of the hot proto-neutron star, and $\rho$ is
the matter density. Using the thermodynamic relation for the entropy
at constant chemical potential, $\mu$,
\begin{equation}
S_{\rm total} = \left( \frac{ \delta P}{\delta T} \right)_{\mu},
\end{equation}
and integrating Eq. (\ref{a1}) we can write entropy per baryon, $S$,
as
\begin{equation}
\label{a2}
T S = \frac{G M_{\rm NS} m_B}{r},
\end{equation}
where $m_B$ is the average mass of one baryon, which we take to be the
nucleon mass.
In the region above the neutron star the material is
radiation dominated and the entropy per baryon can be written in the
relativistic limit as
\begin{equation}
\label{a3}
\frac{S}{k} = \frac{2 \pi^2}{45} \frac{g_s}{\rho_B} \left(
\frac{k T}{\hbar c} \right)^3 ,
\end{equation}
where the statistical weight factor is given by
\begin{equation}
\label{a4}
g_s = \sum_{\rm bosons} g_b + \frac{7}{8} \sum_{\rm fermions} g_f.
\end{equation}
Assuming a constant entropy per baryon, Eqs. (\ref{a2}) and (\ref{a4})
give the baryon density, $\rho_3$, in units of $10^3$ g cm$^{-3}$ as
\begin{equation}
\label{a5}
\rho_3 \sim 38 \left( \frac{g_s}{11/2} \right) \frac{1}{S_{100}^4 r^3_7} ,
\end{equation}
where $S_{100}$ is the entropy per baryon in units of 100 times
Boltzmann`s constant, $r_7$ is the distance from the center in
units of $10^7$ cm, and we assumed that $M_{\rm NS} = 1.4 M_{\odot}$.
Defining $T_9$ to be temperature in units of $10^9$ K, Eq. (\ref{a2})
takes 
the form
\begin{equation}
T_9 S_{100} \sim \frac{2.25}{r_7} . 
\end{equation}

\begin{figure}[h] \begin{center}
\vspace*{2cm}
\includegraphics[scale=0.6]{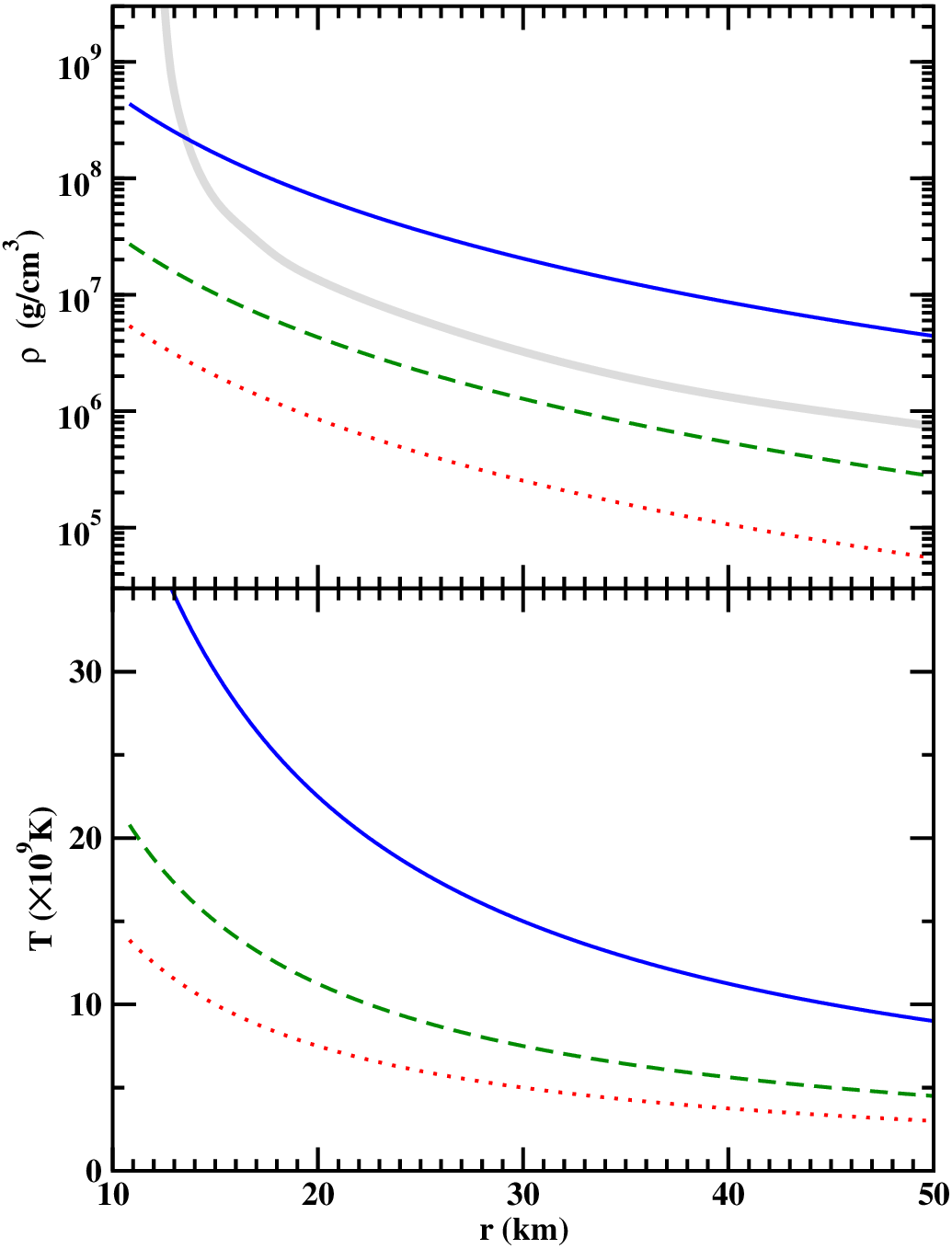}
\vspace*{+0.5cm} \caption{ \label{fig:1}
Solid ($S_{100}$ = 0.5), dashed ($S_{100}$ = 1), and dotted
($S_{100}$=1.5)
lines corresponds to matter density profiles (upper panel), and
temperature
profiles (lower panel) based on heuristic description. The thick band
in the
upper panel is matter density profile from numerical supernova models
for
$t_{PB} \approx 4$ s (taken from ref \cite{Qian:1993dg}).}
\end{center}  \end{figure}

In Figure \ref{fig:1}, we present matter density and temperature
profiles based on heuristic description given in this section. Several
values of $S_{100}$ can be used to describe stages in
the evolution of supernovae. Smaller
entropies per
baryon, $S_{100}\lesssim 0.5$, provide a better description of shock
re-heating
epoch, while larger values,
$S_{100}\gtrsim 1$, describe late times in supernova evolution,
namely, the
neutrino-driven wind epoch. Higher entropy corresponds to less
ordered configurations with smaller baryon densities. In Figure 1 
the statistical
weight factor, $g_s$, is taken to be $11/2$ in the calculation of
matter density since
temperature and entropy per baryon are $T_9 \gtrsim 4$ and $S_{100}
\lesssim 1.5$, respectively.
Under these conditions both photons are electron-positron pairs are
present in the plasma. When
temperature drops, $T_9 \lesssim 4$, only photons present in the
medium and
statistical weight factor has to be taken $g_s \approx 2$.

\subsection{Neutrino Fluxes and Luminosities}

We adopt the prescription in Ref. \cite{Qian:1994wh} for the neutrino
fluxes. If we take the density of particles to be $\rho$, the number
of particles that go through an expanding surface of radius $r$ per 
unit time is
\begin{equation}
\label{2}
\frac{dN}{dt} = \rho \frac{dV}{dt} = 4 \pi r^2 \rho \frac{dr}{dt},
\end{equation}
where $N$ is the total number of particles and $V = 4 \pi r^3$.
Assuming a Fermi-Dirac distribution function for the neutrino
number densities, ignoring neutrino mass in comparison to its energy,
and taking the particle velocity to be $c$, Eq. (\ref{2}) gives for
the flux of neutrinos emitted from the neutrinosphere as 
\begin{eqnarray}
\label{3}
d\phi_{\nu} &=& \frac{d^2 \phi_{\nu}}{dE_{\nu} d \Omega_{\nu}}
dE_{\nu} d \Omega_{\nu}
\nonumber \\
&=& \frac{1}{8\pi^3} \frac{c}{(\hbar c)^3} \frac{E^2_{\nu} dE_{\nu}}{1
+ \exp [ (E_{\nu} - \mu_{\nu}) / kT_{\nu} ]} d \Omega_{\nu},
\end{eqnarray}
where $\mu_{\nu}$ is the neutrino chemical potential. Choosing the
z-direction as the vector that connects the point where the flux is
to be calculated (with radial position $r$) to the center of the
neutron star, one sees
that the azimuthal symmetry still holds, but the polar angle is
bounded by the finite size of the neutrinosphere. Thus
\begin{equation}
\label{4}
\int d \Omega_{\nu} = 2 \pi ( 1 - \cos \theta_0 ),
\end{equation}
where
\begin{equation}
\label{5}
\cos \theta_0  = \sqrt{ 1 - \frac{R^2_{\nu}}{r^2} } \approx
1 - \frac{R^2_{\nu}}{2 r^2}
\end{equation}
with $R_{\nu}$ being the radius of the neutrinosphere.
Within this approximation one then obtains the differential neutrino
flux as
\begin{equation}
\label{6}
\frac{d \phi_{\nu}}{dE_{\nu}} =  \frac{1}{8\pi^2}
\frac{c}{(\hbar c)^3} \frac{R^2_{\nu}}{r^2}
\frac{E^2_{\nu} }{1
+ \exp [ (E_{\nu} - \mu_{\nu}) / kT_{\nu} ]} .
\end{equation}
Using similar reasoning one can write an expression for the
neutrino luminosity. Replacing $N$ in Eq. (\ref{2}) by the total
energy and the matter density $\rho$ by the energy density one
can write down
\begin{equation}
\label{7}
L_{\nu} = 4 \pi r^2 c \frac{1}{2\pi \hbar^2} \int
\frac{E_{\nu} d^3 {\bf p}_{\nu}  }{1
+ \exp [ (E_{\nu} - \mu_{\nu}) / kT_{\nu} ]}.
\end{equation}
Again ignoring the neutrino mass as compared to its energy and
using Eqs. (\ref{4}) and (\ref{5}) to do the angular integration
we obtain
\begin{equation}
\label{8}
L_{\nu} = \frac{c R_{\nu}^2}{ 2 \pi (\hbar c)^3} (k T_{\nu})^4
F_3 (\eta) ,
\end{equation}
where $\eta = \mu_{\nu} / kT_{\nu}$ and $F_3 (\eta)$ is the
relativistic Fermi integral
\begin{equation}
\label{9}
F_3 (\eta) = \int_0^{\infty} \frac{x^3}{1 + \exp ( x - \eta )}
dx .
\end{equation}
In our calculations we take $ \eta =0$ and use the value
$ F_3 (0) = 7 \pi^4 /120$.
Often it is convenient to express the neutrinosphere
radius in terms of the neutrino luminosity. Rewriting $R_{\nu}$
in terms of $L_{\nu}$ and inserting it into Eq. (\ref{6}) we get
\begin{equation}
\label{10}
\frac{d \phi_{\nu}}{dE_{\nu}} =  \frac{1}{4 \pi r^2}
\frac{L_{\nu}}{ (kT_{\nu})^4 F_3 (\eta)}
\left(
\frac{E^2_{\nu} }{1
+ \exp [ (E_{\nu} - \mu_{\nu}) / kT_{\nu} ]} \right) .
\end{equation}

\subsection{Reaction Rates}

The dominant reactions that control the $n/p$ ratio is the capture
reactions on free nucleons
\begin{equation}\label{eq:11}
\nu_e + {\rm n}  \rightleftharpoons {\rm p}+ e^{-} ,
\end{equation}
and
\begin{equation}\label{eq:12}
\bar{\nu}_e + {\rm p} \rightleftharpoons {\rm n} + e^{+} .
\end{equation}
We take the cross sections for the forward reactions to be
\cite{Qian:1995ua}
\begin{equation}\label{eq:13}
\sigma_{\nu_{e}}(E_{\nu_e})\approx 9.6 \times 10^{-44}
\left( \frac{E_{\nu_e}+\Delta_{np}} {\rm MeV}\right)^2 {\rm cm}^2 ,
\end{equation}
and
\begin{equation}\label{eq:14}
\sigma_{\bar{\nu_e}}(E_{\bar{\nu_e}}) \approx 9.6 \times 10^{-44}
\left(\frac{E_{{\bar\nu_e}}-\Delta_{np}} {\rm MeV}\right)^2 {\rm cm}^2
,
\end{equation}
where $\Delta_{np}  \approx 1.293 $ MeV is the neutron proton mass
difference. For simplicity we ignored weak magnetism and recoil
corrections, which may be important \cite{Horowitz:1999fe}. These
corrections cancel for the former cross section, but add for the
latter one.
The rates of these reactions can be written as \cite{Fuller:1981mv}
\begin{equation}
\label{15}
\lambda = \int \sigma (E)_{\nu} \frac{d \phi_{\nu}}{dE_{\nu}}
dE_{\nu} .
\end{equation}

To calculate the neutrino capture rates on nuclei one needs to include
all possible transitions from the parent to the daughter nucleus,
including not only the allowed transitions, but also transitions to
the isobaric analog states, Gamow-Teller resonance states, transitions
into continuum, as well as forbidden transitions. Aspects of such
calculations are discussed in Refs. \cite{Fuller:1995ih} and
\cite{McLaughlin:1995ig} (see also \cite{Volpe:2004dg}).
Careful input of such reaction rates in supernova simulations is
especially crucial to assess the possibility of core-collapse
supernovae as a site of the 
r-process nucleosynthesis \cite{Balantekin:2003ip,Meyer:1998sn}.

Because of their charged-current interaction electron neutrinos 
may play a role in reheating the stalled shock as well as 
regulating the neutron-to-proton ratio. In contrast, since the
energies of the muon and tau neutrinos and antineutrinos produced are
too low to produce charged leptons, these neutrinos interact only with
the  
neutral-current interactions. Recently significant attention was 
directed towards understanding the $\nu_{\mu}$ and $\nu_{\tau}$ 
spectra formation. Neutrinos remain in local thermal equilibrium 
as long as they can participate in reactions that allow exchange 
of energy and neutrino pair creation or annihilation. It turns 
out that the neutrino bremsstrahlung process
\begin{equation}
\label{new1}
N + N \leftrightarrow N + N + \nu + \bar{\nu}
\end{equation}
is more effective than the annihilation process $\nu + \bar{\nu} 
\leftrightarrow e^+ + e^-$ at equilibrating neutrino number 
density \cite{suzuki}. (However the neutrino-neutrino 
annihilation process $\nu_e + \bar{\nu}_e  
\leftrightarrow \nu_{\mu} + \bar{\nu}_{\mu}$ is one of the 
primary sources of muon and tau neutrinos 
\cite{Buras:2002wt}).
The impact of the neutrino bremsstrahlung 
process on equilibrating the energy spectra seems to be 
comparable to that of
\begin{equation}
\label{new2}
\nu_{\mu,\tau} + e^- \rightarrow \nu_{\mu,\tau} + e^-. 
\end{equation}
The most effective process to exchange energy is 
\cite{Hannestad:1997gc} 
\begin{equation}
\label{new3}
\nu + N + N \rightarrow N + N + \nu ,
\end{equation}
which dominates the neutrino spectra formation. Finally 
recoil corrections to the $\nu N$ interactions are very 
important in the formation of the $\nu_{\mu}$ and 
$\nu_{\tau}$ spectra as they permit energy exchange 
\cite{Raffelt:2001kv}.

\subsection{Electron Fraction}

The electron fraction, $Y_e$, is the net number of electrons (number
of electrons minus the number of positrons) per baryon:
\begin{equation}
\label{16}
Y_e = ( n_{e^-} - n_{e^+} ) / n_B ,
\end{equation}
where $n_{e^-}$, $n_{e^+}$, and $n_B$ are number densities of
electrons, positrons, and baryons, respectively. Introducing $N_j$,
number of species of kind $j$ per unit volume, and $A_j$, atomic
weight of the $j$-th species, one can write down expressions for the 
mass fraction, $X_j$
\begin{equation}
\label{17}
X_j = \frac{N_j A_j}{ \sum_i  N_i A_i},
\end{equation}
and the number abundance relative to baryons, $Y_j$, 
\begin{equation}
\label{18}
Y_j = \frac{X_j}{A_j} = \frac{N_j}{\sum_i N_i A_i} .
\end{equation}
The electron fraction defined in Eq. (\ref{16}) can then be rewritten
as
\begin{eqnarray}
\label{19}
Y_e &=& \sum_i Z_i Y_i = \sum_i \left( \frac{Z_i}{A_i} \right) X_i
\nonumber \\
&=& X_p + \frac{1}{2} X_{\alpha} +  \sum_h \left( \frac{Z_h}{A_h}
\right) X_h ,
\end{eqnarray}
where $Z_i$ is the charge of the species of kind $i$, and the mass
fractions of protons, $X_p$, alpha particles, $X_{\alpha}$, and
heavier nuclei (``metals''), $X_h$, are explicitly indicated.

The rate of change of the number of protons can be expressed as
\begin{equation}
\label{20}
\frac{dN_p}{dt} = - ( \lambda_{\bar{\nu}_e} + \lambda_{e^-} ) N_p
+  ( \lambda_{\nu_e} + \lambda_{e^+} ) N_n ,
\end{equation}
where $\lambda_{\nu_e}$ and $\lambda_{e^-}$ are the rates of the
forward and backward reactions in Eq. (\ref{eq:11}) and
$\lambda_{\bar{\nu}_e}$ and $\lambda_{e^+}$ are the rates of the
forward and backward reactions in Eq. (\ref{eq:12}). Since the
quantity $\sum_i N_i A_i$ does not change with neutrino
interactions, one can
rewrite Eq. (\ref{20}) in terms of mass fractions
\begin{equation}
\label{21}
\frac{dX_p}{dt} = - ( \lambda_{\bar{\nu}_e} + \lambda_{e^-} ) X_p
+  ( \lambda_{\nu_e} + \lambda_{e^+} ) X_n .
\end{equation}
In the hot bubble the rates are usually expressed in terms of the
radial velocity field, $v(r)$, above the neutron star, i.e. $dY/dt =
v(r) [dY/dr]$. A careful study of the influence of nuclear composition
on $Y_e$ in the post-core bounce supernova environment is given in
Ref. \cite{McLaughlin:1997qi} to which we refer the reader for further
details.

If no heavy nuclei are present we can write
\begin{equation}
\label{22}
Y_e = X_p + \frac{1}{2} X_{\alpha} .
\end{equation}
Because of the very large binding
energy, the rate of alpha particle interactions with
neutrinos is nearly zero and we can write $dY_e /dt = dX_p /dt$.
Using the constraint $X_p + X_n + X_{\alpha} = 1$ and Eq.
(\ref{22}), Eq. (\ref{21}) can be rewritten as
\begin{equation}
\label{23}
\frac{dY_e}{dt} = \lambda_n - ( \lambda_p + \lambda_n) Y_e
+ \frac{1}{2} ( \lambda_p - \lambda_n ) X_{\alpha},
\end{equation}
where we introduced the total proton loss rate $\lambda_p =
\lambda_{\bar{\nu}_e} + \lambda_{e^-}$ and the total neutron loss
rate $\lambda_n = \lambda_{\nu_e} + \lambda_{e^+}$. It has been
shown that when the rates of these processes are rapid as compared
to the outflow rate a ``weak chemical equilibrium'' is established
\cite{Qian:1993dg}. The weak freeze-out radius is defined to be
where the neutron-to-proton conversion rate is less than the
outflow rate of the material. If the plasma reaches a weak
equilibrium stage then $Y_e$ is no longer changing: $dY_e /dt =
0$. From Eq. (\ref{23}) one can write the equilibrium value of the
electron fraction
\begin{equation}
\label{24}
Y_e = \frac{\lambda_n}{\lambda_p + \lambda_n}
+ \frac{1}{2} \frac{\lambda_p - \lambda_n}{\lambda_p + \lambda_n}
X_{\alpha} .
\end{equation}

Different flavors of neutrinos decouple at different radii. Since
$\nu_{\mu}$ and $\nu_{\tau}$ (and their antiparticles) interact with
the ordinary matter only with the neutral current interactions, they
decouple deeper in the core 
and have a large average energy. Electron neutrinos and
antineutrinos have additional charged-current interactions with
neutrons and protons respectively. Since in the supernova environment
there are more neutrons, electron antineutrinos decouple after
$\nu_{\mu}$'s and $\nu_{\tau}$'s, but before electron
neutrinos. Consequently one has a hierarchy of average neutrino
energies:
\begin{equation}
\langle E_{\nu_e} \rangle \leq  \langle E_{\bar{\nu}_e} \rangle  \leq
\langle E_{\nu_x, \bar{\nu}_x} \rangle ,
\end{equation}
where $\nu_x$ stands for any combination of $\nu_{\mu}$'s and
$\nu_{\tau}$'s. However a more complete description of the
microphysics suggests that this hierarchy of average energies may not
be very pronounced \cite{Raffelt:2001kv,Buras:2002wt,Keil:2002in}.
This microphysics is dominated by the inelastic neutrino-nucleon 
interactions discussed in Section 2.3. 

\begin{figure}[t] \begin{center}
\vspace*{+1.5cm}
\includegraphics[scale=0.6]{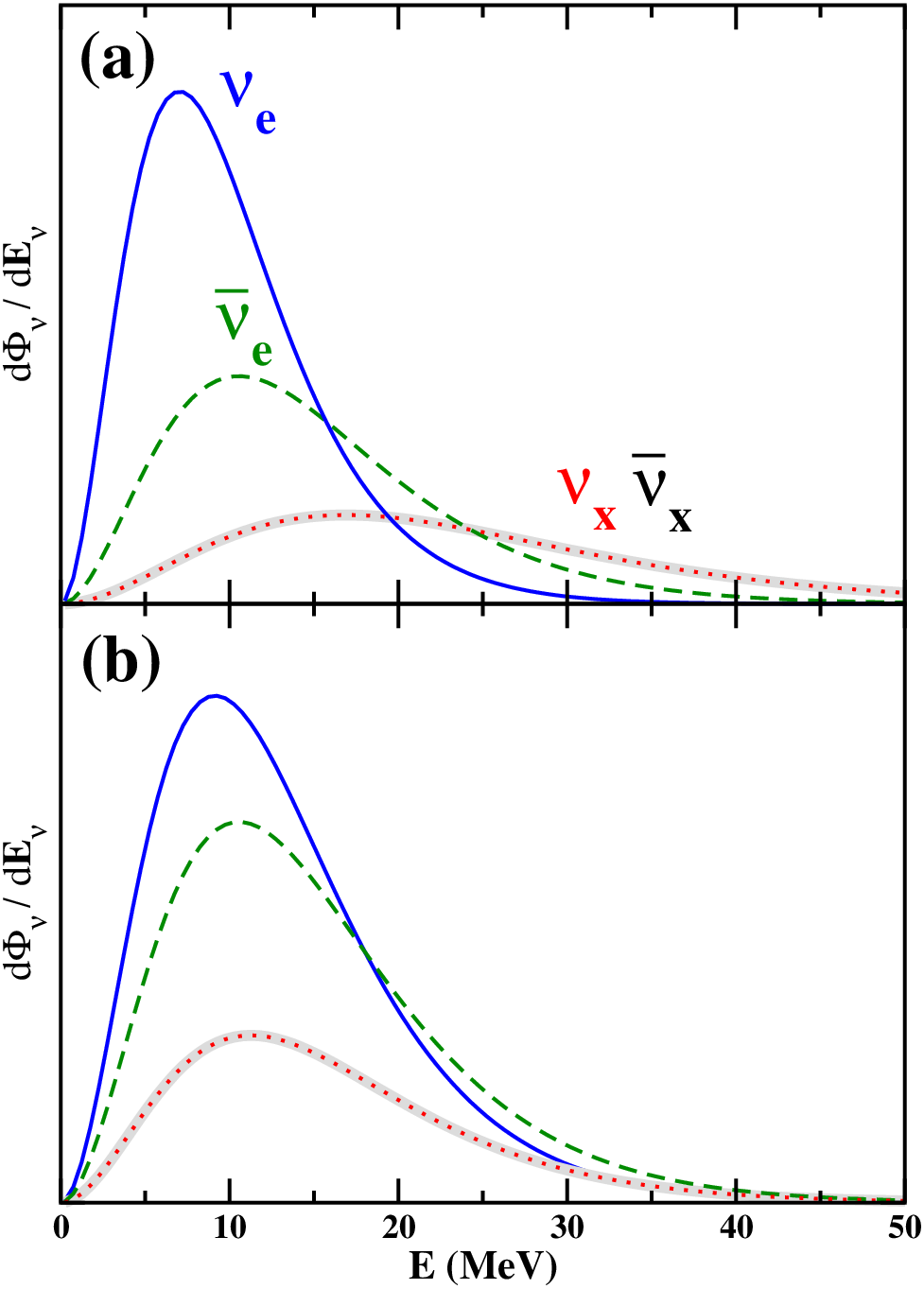}
\vspace*{+0cm} \caption{ \label{fig:2}
Initial 
differential neutrino fluxes in arbitrary units. Solid, dashed, 
dotted, and thick 
lines correspond to the distributions of $\nu_e$, $\bar{\nu}_e$, 
$\nu_{x}$, and  $\bar{\nu}_{x}$.  
a) Fluxes with $\langle E_{\nu_e} \rangle = 10 $
MeV,
$\langle E_{\bar \nu_{e}} \rangle =15 $ MeV, and
$\langle E_{\nu_{x},\bar\nu_{x}} \rangle =24 $ MeV. Neutrino and
antineutrino luminosities are taken to be equal 
for all flavors. Distributions of
$\nu_{x}$,$\bar{\nu}_{x}$
have much longer tails up to 100 MeV not shown in the figure.
b)  Fluxes with $\langle E_{\nu_e} \rangle =13 $
MeV,
$\langle E_{\bar \nu_{e}} \rangle =15 $ MeV, and
$\langle E_{\nu_{x},\bar\nu_{x}} \rangle = 16 $ MeV.
$\nu_e$ and $\bar{\nu}_e$ luminosities are taken to be equal. 
Luminosities of other flavors are taken to be 
half of the $\nu_e$ and $\bar{\nu}_e$ luminosities.}
\end{center}  \end{figure}

In Figure \ref{fig:2}, we present initial differential neutrino
fluxes. In the upper panel the typical post-bounce neutrino energies 
are taken as the representative values of $\langle E_{\nu_e} 
\rangle =10 $ MeV, $\langle E_{\bar \nu_{e}} \rangle =15 $ MeV, and
$\langle E_{\nu_{x},\bar\nu_{x}} \rangle =24 $ MeV.
(Average neutrino temperature for each flavor
can be calculated using the relation: $T_{\nu}=\langle E_{\nu}
\rangle/3.151$). Here neutrino and
antineutrino luminosities are taken to be equal for all flavors.
We express the neutrino luminosities in more convenient units of
$10^{51}\,{\rm ergs}\,{\rm s}^{-1} $. Equal luminosities of
$L^{51}_{\nu} =1$ is typically a good approximation for
the neutrino-driven wind epoch.
However, at earlier epochs, $\nu_e$ and  $\bar{\nu}_e$ luminosities
can be as
large as $L^{51}_{\nu} =10$. Except, through the neutronization burst,
for few
milliseconds ${\nu}_e$ luminosity can reach values of $L^{51}_{\nu}
=100$, an order of magnitude larger than $\bar{\nu}_e$ during the same
period. In the lower panel of Figure 2 we 
examine a less-pronounced hierarchy of neutrino energies. 
Here we adopt the representative values of $\langle E_{\nu_e} 
\rangle =13 $ MeV, $\langle E_{\bar \nu_{e}} \rangle = 15 $ MeV, and
$\langle E_{\nu_{x},\bar\nu_{x}} \rangle = 16 $ MeV. Here the
luminosities of $\nu_x$ and $\bar{\nu}_x$ are taken to be the half of
the value of the equal luminosities of $\nu_e$ and $\bar{\nu}_e$.  

In our calculations in Section 4, we adapt these initial
distributions of  $\nu_e$,
$\bar\nu_e$,   $\nu_x$  and $\bar\nu_x$ at neutrinosphere,
$R_{\nu}$=10 km,
with the indicated luminosities and follow the evolution of
differential neutrino fluxes (number of neutrinos 
per unit energy per unit volume).

\subsection{Alpha Effect}

At high temperatures alpha particles are absent and the second
term in Eq. (\ref{24}) can be dropped. In the region just below
where the alpha particles are formed approximately one second
after the bounce, the temperature is less then $\sim 1$ MeV. Here
both the electron and positron capture rates are very small and
$Y_e$ can be approximated as
\begin{equation}
\label{25}
Y_e^{(0)} = \frac{1}{1 + \lambda_{\bar{\nu}_e} /
\lambda_{\nu_e}} .
\end{equation}

As the alpha particle mass fraction increases (when $T_9$ drops below
8) free nucleons
get bound in alphas and, because of the large binding energy of
the alpha particle, cease interacting with neutrinos.
This phenomenon is called
``alpha effect'' \cite{Fuller:1995ih}. Using Eq. (\ref{25}) one
can rewrite Eq. (\ref{24}) as
\begin{equation}
\label{26}
Y_e = Y_e^{(0)} + \left( \frac{1}{2} - Y_e^{(0)} \right)
X_{\alpha} .
\end{equation}
Hence if the initial electron fraction is small ($Y_e^{(0)}
< 1/2$) the alpha effect increases the value of $Y_e$.
Since higher $Y_e$ implies fewer free neutrons, the alpha-effect
negatively impacts r-process nucleosynthesis \cite{Meyer:1998sn}.

Neutrino oscillations, since they can swap energies of
different flavors, can effect the energy-dependent rates in
Eqs. (\ref{25}) and (\ref{26}), changing the electron
fraction. Indeed, transformations between active flavors heat up
$\nu_e$'s and increase $\lambda_{\nu_e}$, driving the electron
fraction to rather large values. Consequently a very large mixing
between active neutrino flavors would have prohibited 
r-process nucleosynthesis in
a core-collapse supernova \cite{Qian:1993dg,Qian:1994wh}. 

Actually electron neutrinos radiated from the proto-neutron stars are
just too energetic to prevent the alpha effect in most cases. One
possibility 
to reduce $\lambda_{\nu_e}$ is to convert active electron neutrinos
into sterile ones which do not contribute to this rate. This
possibility is explored in references \cite{McLaughlin:1999pd},
\cite{Caldwell:1999zk} and \cite{Fetter:2002xx}.


\section{Neutrino Mixing}

Neutrino interactions in matter is a rich subject (for a brief review
see Ref. \cite{Balantekin:1998yb}). While neutrinos are produced through 
weak interactions in flavor eigenstates, they propagate in mass
eigenstates. Mixing angles correspond to rotations describing unitary 
connection
between two bases: 
\begin{equation}
\label{eq:h1}
\left( \begin{array}{c} \nu_e \\ \nu_\mu \\ \nu_\tau \end{array} \right)
   =  U_{\alpha i}
\left( \begin{array}{c} \nu_1 \\ e^{i \phi_1} \nu_2 \\ e^{i \phi_2} \nu_3
\end{array} \right).
\end{equation}  
In our discussion, we follow the notation in Ref. \cite{Balantekin:2003dc}
and denote the neutrino mixing matrix by $U_{\alpha i}$ where $\alpha$ denotes
the flavor index and $i$ denotes the mass index:
\begin{equation}
\label{mixing}
U_{\alpha i} = \left(\matrix{
     1 & 0 & 0  \cr
     0 &  C_{23}  & S_{23} \cr
     0 & - S_{23} &  C_{23} }\right)
 \left(\matrix{
     C_{13} & 0 &  S_{13}^{\ast} \cr
     0 &  1 & 0 \cr
     - S_{13} & 0&  C_{13} }\right)
 \left(\matrix{
     C_{12} & S_{12} &0 \cr
     - S_{12} & C_{12} & 0 \cr
     0 & 0&  1 }\right) .
\end{equation}
In Eq. (\ref{mixing}) $C_{13}$, etc. is the short-hand notation for 
$\cos {\theta_{13}}$, etc.
The notation $S_{13}^{\ast}$ was used to indicate 
$(\sin{\theta_{13}})e^{i\phi}$
where $\phi$ is a CP-violating phase. We will ignore this phase in our
discussion.


We have compelling evidence supporting 
non-zero neutrino 
masses and mixings. 
Two-flavor solar neutrino solution corresponding to $\theta_{12} \sim 
\pi/6$ and 
$\delta m_{12}^2 \sim 8 \times 10^{-5}$ eV$^2$ was identified using
the recent Sudbury Neutrino Observatory (SNO) results (cf.
Refs. \cite{Ahmed:2003kj}, \cite{Balantekin:2003jm}). Measurement of
anti-neutrinos from nuclear power reactors in Japan by the KamLAND experiment 
confirmed this solution and improved the limits on the solar mass square 
difference, $\delta m_{12}^2$, significantly \cite{Araki:2004mb}. 
While it is known that
the solar neutrino mixing angle is large, but not maximal, the atmospheric
mixing angle is large and could be maximal, $\theta_{23} \sim \pi/4$, as 
shown by the 
Super-Kamiokande experiment \cite{Ashie:2004mr}. The latter angle 
is consistent with the 
KEK-to-Kamioka oscillation experiment, K2K \cite{Ahn:2002up}. The
corresponding atmospheric mass square difference is $\delta m_{23}^2
\sim 3\times 10^{-3}$ eV$^2$.

The size of the last mixing angle, $\theta_{13}$ is currently best limited 
by the combined CHOOZ \cite{Apollonio:2002gd} and Palo Verde
\cite{Boehm:2001ik} reactor experiments and SK atmospheric 
data. This is due to the null results from the reactor $\bar{\nu}_e$
disappearance over the $\delta m_{23}^2$ distance scale. The upper 
bound on this
angle from KamLAND and the 
solar neutrino data gets stronger (especially in the
region with small atmospheric mass square difference where CHOOZ 
reactor bound is
relatively weak), and even dominates, as this data get refined 
\cite{Balantekin:2004hi}. Both measurements of the width of the Z boson and
oscillation interpretation of the neutrino data, with the notable exception 
of the LSND signal \cite{Aguilar:2001ty}, 
favor three
generations of light active neutrino species. If LSND is confirmed, the most
likely explanation could be the existence of a fourth neutrino (a relatively 
heavier
sterile neutrino) since agreement between KamLAND and combined solar 
experiments
already disfavors the interpretation of the LSND anomaly 
with CP violation.  

In the region above the supernova core density is still high, but steeply 
decreases. Matter-enhanced oscillations 
mediated by the solar mass square difference are impossible in the
region close to the core which is suitable for the 
r-process. However, at late neutrino
driven epoch baryon density could be low enough to allow resonances through
$\delta m_{13}^2$, which is comparable to the 
atmospheric mass square difference. At such late times neutrino flux is  
expected to be low. We examine prospects of r-process at such environments 
in the next section.


We describe neutrino mixing within the density matrix formalism
\cite{Qian:1994wh,Sigl:1992fn,Raffelt:1992uj,Loreti:1994ry}. We 
assume that the electron neutrino mixes
with a linear combination of $\mu$ and $\tau$ neutrinos. The neutrino
oscillations are 
mediated by  the matter mixing angle for transformation between
1st and 3nd mass eigenstates, $\theta_{13}$, and the corresponding atmospheric 
mass
square difference, $\delta m_{13}^2 $. The mixing between $\mu$ and $\tau$
neutrino flavors does not have any significant effect on our results as
long as total luminosities and corresponding average energies are equal
and their mixing is maximal. In this limit, mixing between 2nd and 3nd mass
eigenstates, can be rotated away and effectively electron neutrinos
oscillates into some linear combination of  $\mu$ and $\tau$ neutrinos 
\cite{Balantekin:1999dx}. 

In the rest of the paper $\theta$ and
$\delta m^2$ refer to $\theta_{13}$ and $\delta m_{13}^2$.
This assumption simplifies the discussion and allows us to write the
two-flavor density matrices as
\begin{equation}\label{eq:h2}
\rho =  \left( \begin{array}{cc}
\rho_{ee} & \rho_{ex} \\ \rho_{xe} & \rho_{xx} \end{array} \right)
=\frac{1}{2} \left( P_0 + {\bf P} \cdot {\bf \sigma} \right)
\end{equation}
and
\begin{equation}\label{eq:h3}
\bar\rho =  \left( \begin{array}{cc}
\bar\rho_{ee} & \bar\rho_{ex} \\ \bar\rho_{xe} & \bar\rho_{xx}
\end{array}
\right)
=\frac{1}{2} \left( \bar P_0 + \bar{\bf P} \cdot {\bf \sigma}
\right),
\end{equation}
where we introduced the polarization vectors for neutrinos and
anti-neutrinos, 
${\mathbf P}_p$ and $\bar{{\mathbf P}}_p$. The diagonal elements in these
expressions are initially given
by the expression in Eq. (\ref{10}). Non-diagonal elements are
initially zero but may become non-zero during the neutrino evolution.

Equations governing the evolution of neutrinos and antineutrinos
can be cast into the forms \cite{Pastor:2002we}
\begin{equation}\label{eq:h4a}
\partial_r{\mathbf P}_p = \left\{
+{\mathbf \Delta}_p + \sqrt2\,G_{\rm F} \left[ N_{e} \hat{\bf z}
+ \int d{\bf q} \left( 1 - \frac{\bf p . q}{p \> q} 
\right)
({\mathbf P_{\bf p}}- {\mathbf{\overline P}}_{\bf q}) \right]  
\right\}
\times 
{\mathbf P}_p ,
\end{equation}
and
\begin{equation}\label{eq:h5a}
\nonumber \partial_r{\mathbf{\overline P}}_p = \left\{
-{\mathbf \Delta}_p + \sqrt2\,G_{\rm F} \left[ N_{e} \hat{\bf z}
+  \int d{\bf q} \left( 1 - \frac{\bf p . q}{p \> q} 
\right) ({\mathbf P_{\bf p}}- {\mathbf{\overline P}}_{\bf q}) \right]  
\right\}
\times {\mathbf{\overline P}}_p , 
\end{equation}
where 
\begin{equation}\label{eq:b5}
{\mathbf \Delta}_p = \frac{\delta m^2}{2p}(\sin 2\theta
{\mathbf {\hat x}} - \cos 2 \theta {\mathbf {\hat z}}),
\end{equation}
Integrating Eqs. (\ref{eq:h4a}) and (\ref{eq:h5a}) exactly 
in the supernova 
environment is, at the moment, an unsolved problem. Indeed the exact 
solutions of these coupled, 
non-linear differential equations are expected to be very complicated. 
Instead we adopt the approximation proposed in Ref. \cite{Qian:1994wh}
and also adopted in Ref. \cite{Pastor:2002we}. In this approximation 
one uses flux-averaged values to obtain 
\begin{equation}\label{eq:h4}
\partial_r{\mathbf P}_p = \left(
+{\mathbf \Delta}_p +\sqrt2\,G_{\rm F} N_{e} \hat{\bf z}
+ \sqrt2 G_{\rm F} F(r)
({\mathbf J}- {\mathbf{\overline J}}) \right) \times {\mathbf P}_p
\end{equation}
and
\begin{equation}\label{eq:h5}
\nonumber \partial_r{\mathbf{\overline P}}_p = \left(
-{\mathbf \Delta}_p + \sqrt2\,G_{\rm F} N_{e} \hat{\bf z}
+ \sqrt2 G_{\rm F} F(r)
({\mathbf J}- {\mathbf{\overline J}}) \right)
\times {\mathbf{\overline P}}_p
\end{equation}
In these equations
${\mathbf J}$ is the polarization integrated over all momentum modes,
and
$F(r)=\frac{1}{2}[1-(1-R_\nu^2/r^2)^{1/2}]$ is the geometrical factor 
introduced earlier (cf. Eqs. (\ref{4}) and (\ref{5})). 

\begin{figure}[t] \begin{center}
\vspace*{+1.5cm}
\includegraphics[scale=0.6]{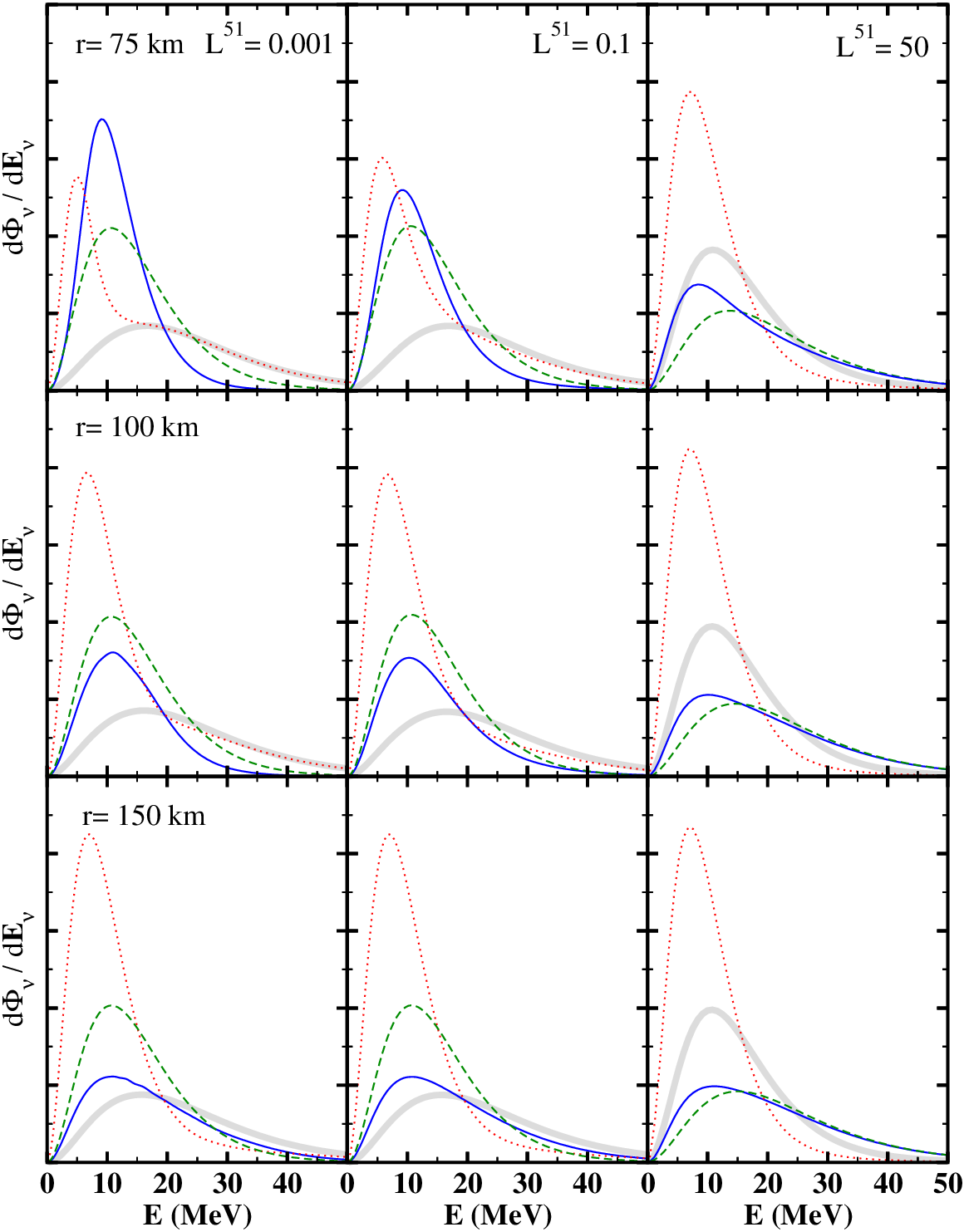}
\vspace*{+0cm} \caption{ \label{fig:3}
Evolution of differential neutrino fluxes in arbitrary units. Solid, dashed, 
dotted, and thick lines correspond to the distributions of $\nu_e$, 
$\bar{\nu}_e$
$\nu_\mu$, and  $\bar{\nu}_\mu$. In each panel, $1/r^2$ dependence is removed.
Columns are for $L^{51}$ = 0.001, 0.1, and 50 from left to the right 
corresponding to very weak, moderate, 
and very strong neutrino self interaction contributions to the evolution. Rows
are calculated at 
$r =$ 75 km, 100 km, and 150 km showing the evolution of distributions
along neutrino path. See text for details.}
\end{center}  \end{figure}
 

\section{Results and Discussion}

In our calculations, we concentrate on the late neutrino-driven wind epoch
which is expected to have larger entropy. We adopt $S_{100}$ = 1.5 and 
the $g_S$ = 2, since in this epoch the
temperature is too low at later times 
to have electron-positron pairs to be adequately 
represented.

\begin{figure}[h] \begin{center}
\vspace*{+1.5cm}
\includegraphics[scale=0.45]{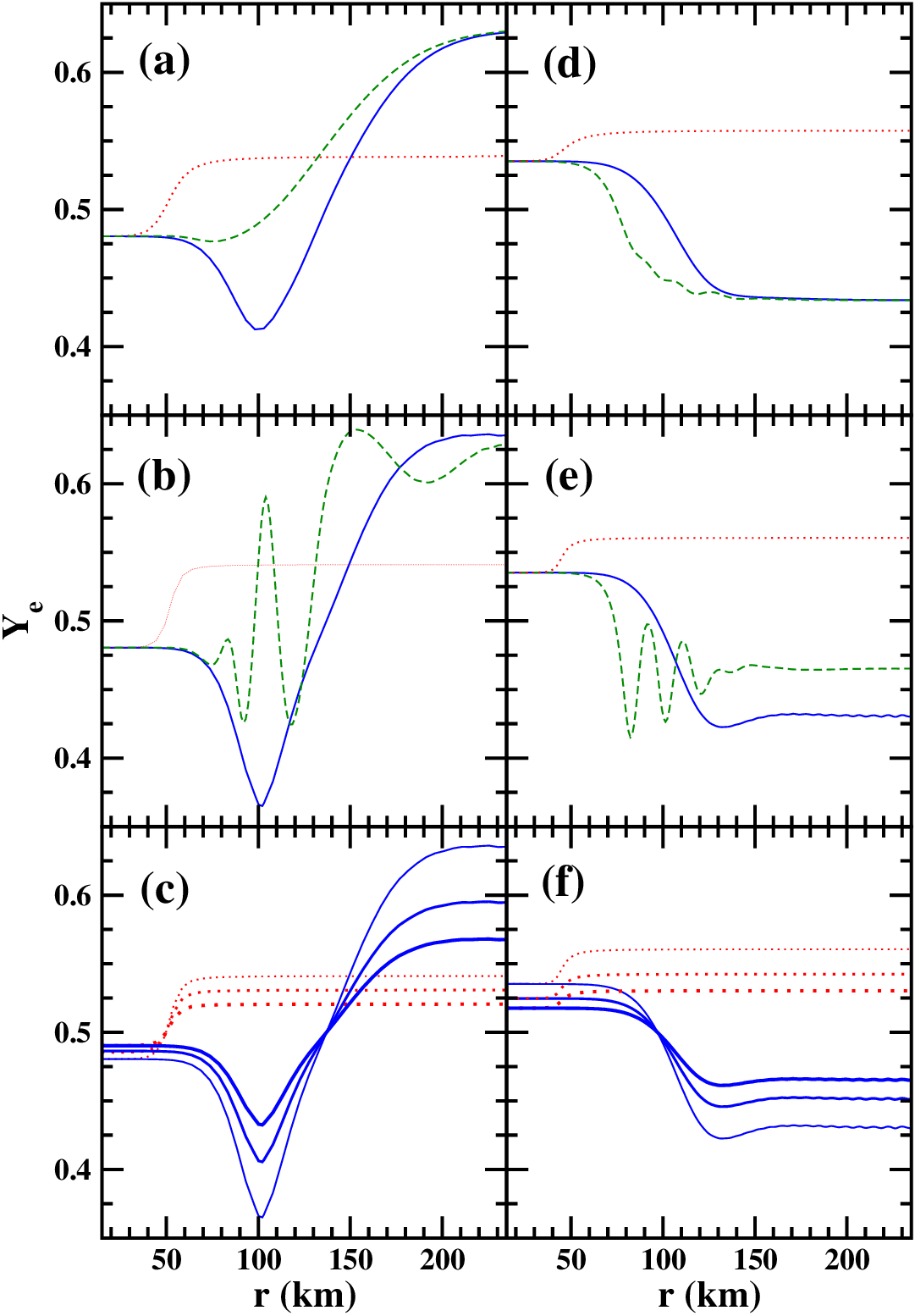}
\vspace*{+0cm} \caption{ \label{fig:4}
Initial neutrino fluxes and
luminosities are taken to be those in Figure 2(a) in the left column 
and those in Figure 2(b) in the right column. 
$S_{100}$ is taken to be 1.5. 
In panels (a) and
(b), solid, dashed and dotted lines correspond to the equal 
luminosities of 
$L^{51}$ = 0.001, 0.1, and 50 for all flavors, 
indicating very weak, moderate, and very-strong neutrino
self interaction contributions to the evolution. 
In panels (d), and (e) $\nu_e$ and $\bar{\nu}_e$ luminosities are
taken to be $L^{51}$ = 0.002, 0.2, and 200 (solid, dashed and dotted
lines, respectively). In (d) and (e) the luminosities of other
flavors are taken to be $L^{51}$ = 0.001, 0.1, and 100 (solid, dashed
and dotted lines, respectively).  
\textbf{(a, d)}
Equilibrium electron fraction as a function of the distance from the core with
mixing parameters $\theta_{13} \sim \pi/10$ and $\delta m_{13}^2 \sim 3 \times
10^{-3}$ eV$^2$. 
\textbf{(b, e)}
Equilibrium $Y_e$ as a function of the distance from the core with
mixing parameters $\theta_{13} \sim \pi/20$ and $\delta m_{13}^2 \sim 3 \times
10^{-3}$ eV$^2$. 
\textbf{(c)}
Same as (b) but the impact of alpha particle formation 
is included according to the Eq. (\ref{24}).  
$Y_e$ is shown when $X_{\alpha}$=0, 0.3 and 0.5 
(thin, medium and thick lines) 
for $L^{51}$=0.001 and 50 (solid and dotted sets of lines).
\textbf{(f)}
Same as (e) but the impact of alpha particle formation 
is included as in (c).  
$Y_e$ is shown when $X_{\alpha}$=0, 0.3 and 0.5 
(thin, medium and thick lines) 
for $L_{\nu_e}^{51}$=0.002 and 200 (solid and dotted sets of lines).}
\end{center}  \end{figure}

In Figure \ref{fig:3} we present differential neutrino
fluxes at several stages of the evolution in order to provide a 
better understanding
of this mechanism. The mixing parameters are chosen as  $\theta\sim
\pi/10$ and $\delta m^2 \sim 3 \times 10^{-3}$ eV$^2$. Solid, dashed, 
dotted, and thick lines correspond to the distributions 
of $\nu_e$, $\bar{\nu}_e$, $\nu_x$,and
$\bar{\nu}_x$. In each panel, $1/r^2$ dependence is removed.
For this figure the initial distributions are taken as 
the values given in Figure 2a). 
Columns are for $L^{51}$ = 0.001, 0.1, and 50 from left to the 
right corresponding to very weak, moderate, 
and very strong neutrino self interaction contributions to the
evolution. Rows 
show neutrino flux at $r = $ 75 km, 100 km, and 150 km, 
exhibiting the evolution of neutrino distributions
along the neutrino path. These should be compared to the initial (at
$\sim 10$ 
km) neutrino distributions given in Figure \ref{fig:2}a). 
Electron fractions corresponding to these 
luminosities are given in the left column of Figure \ref{fig:4}.  

$\bullet$ \textbf{First column of Figure \ref{fig:3}, $L^{51}$ = 0.001:} 
Initially differential neutrino fluxes for $\nu_e$ and $\nu_x$ are the same 
at $\sim$ 20 MeV as seen in Figure \ref{fig:2}a). At energies lower than 20 
MeV, differential flux of $\nu_e$ is higher than that of $\nu_x$ whereas 
at energies higher than 20 MeV the numbers reverse. 
The transformation starts at the low 
energy tail of the distributions. As we gradually move to 
lower densities resonance moves up to higher energies. 
Until we reach the resonance at 20 MeV the luminosity of $\nu_e$'s 
decreases.   
At r = 100 km electron neutrinos below 20 MeV are mostly swapped with 
$\nu_x$'s while above 20 MeV there is yet no significant
transformation. 
This point corresponds to the minimum of $\lambda$ of Eq. (\ref{15})
for neutrinos (after the $1/r^2$-dependence is taken out).
Since antineutrinos are not transformed $\lambda$ is constant for
them. As a results, r $\sim$ 100 km 
corresponds to a dip in $Y_e$ (solid line in figure
\ref{fig:4}\textbf{(a)}). 
Since above 20 MeV the initial differential flux of $\nu_e$'s is 
smaller than that of $\nu_x$'s, transformations after r $\sim$ 100 km would 
increase the number of high energy $\nu_e$'s.    
 As neutrinos travel to
regions further away (r= 150 km, where baryon density is much lower),
swapping of high energy tail of the $\nu_e$ and $\nu_x$ 
distributions is completed, 
and $Y_e$ approaches to the asymptotic fixed value. 

$\bullet$ \textbf{Second column of Figure \ref{fig:3}, $L^{51}$ = 0.1:} 
This column describes the 
same evolution of neutrinos except that neutrino self-interaction effects 
play a more pronounced role. The chosen value of the luminosity 
could be representative of such late times in neutrino-driven wind epoch.
Because of the small contributions of the self interactions, resonance
region 
is relatively wider. No dip is observed in $Y_e$ (dashed line in figure
\ref{fig:4}\textbf{(a)}) at r = 100 km because transformation of low and high
energy ends of the distributions happen almost simultaneously. 
There is also small transformation 
between $\bar{\nu}_e$'s and $\bar{\nu}_x$'s.

$\bullet$ \textbf{Third column of Figure \ref{fig:3}, $L^{51}$ = 50:} 
This is an extreme case to
illustrate the limit at which neutrino self interactions dominate.  
Swapping of both neutrinos and
antineutrinos occur and $Y_e$ reaches its equilibrium value rapidly. 
After the transformation, electron neutrinos and
antineutrinos assume similar luminosities and distributions since they
are swapped with $\nu_x$'s and $\bar{\nu}_x$'s. Note that this
equilibrium value is again different from 0.5 because of the threshold 
effects (due to the neutron-proton mass difference) in the reaction cross 
sections (cf. Eqs. (\ref{eq:13}) and (\ref{eq:14})).  

In the left column of Figure 4 initial neutrino fluxes and
luminosities are taken to be those in Figure 2(a) and in the right
column initial neutrino fluxes and
luminosities are taken to be those in Figure 2(b).  
In Figure \ref{fig:4}(a) 
we present the equilibrium
electron fraction, $Y_e$, as a function of the 
distance from the core for the three different
cases of neutrino flux, $L^{51}$ = 0.001, 0.1, and 50, 
each corresponding to one of the columns  
in Figure \ref{fig:3}.
In  Figure \ref{fig:4}(a) we use the same neutrino parameters as in 
Figure \ref{fig:3}. 
One could argue that the relatively large mixing angle of $\pi / 10$ is 
already disfavored by
CHOOZ. To explore the implications of a smaller mixing angle, 
in Figure \ref{fig:4}(b) we show electron fractions calculated 
with the more realistic mixing parameters $\theta\sim
\pi/20$ and $\delta m^2 \sim 3 \times 10^{-3}$ eV$^2$.  
The dip in $Y_e$ plot is sharper since
resonance region will be much narrower with the smaller mixing angle.

In Figures \ref{fig:4} a) and b), to calculate the equilibrium 
electron fraction, we ignored possible effects of the alpha particles and 
used Eq. (\ref{25}) to evaluate $Y_e$. In \ref{fig:4}(c), we 
explore possible effects of alpha particles by using Eq. (\ref{24}) to 
calculate $Y_e$ for three different values of $X_{\alpha}$. The 
``alpha effect'' is manifest: as $X_{\alpha}$ gets larger, it pulls the 
value of the electron fraction closer to 0.5. 

In the right column of Figure 4 initial neutrino fluxes and
luminosities are taken as the generic ``almost equal'' energies and
different luminosities case shown in Figure 2(b). It is immediately
obvious that the situation is markedly different in this case. Even 
though the initial
$\nu_e$ and $\bar{\nu}_e$ spectra are about the same (cf. Figure
2(b)), $Y_e$ initially is greater than $0.5$ because of the effect of
the neutron-proton mass difference in Eqs. (\ref{eq:13} and
(\ref{eq:14}). When the effects of the neutrino self-interaction terms
are minimal (the low-luminosity case indicated by the solid lines in
(d) and (e)), after $\nu_e$'s and $\nu_x$'s swap $\nu_e$ luminosity
decreases significantly causing a big drop in the value of $Y_e$. When
the effects of the self-interaction terms are dominant (the high
luminosity cases indicated by the dotted lines in (d) and (e)), all
flavors swap and the electron fraction eventually reaches to 
the asymptotic values.
The impact of alpha formation, when the hierarchy of
average neutrino energies is less pronounced, is presented in Figure
4(f).

\section{Conclusions}                                          

In this article we investigated conditions for the r-process 
nucleosynthesis at late 
neutrino-driven wind epoch in a core-collapse supernova. We considered
the  
region where the matter-enhanced neutrino transformation is driven by 
$\delta m_{13}^2$, which we took to be comparable to the mass difference 
observed in the atmospheric neutrino oscillations. 

We found that, when
initial luminosities are taken to be equal with a pronounced
hierarchy of neutrino energies, the asymptotic 
value (at large distances from the core) 
of the electron fraction always exceeds 0.5, hindering the r-process 
nucleosynthesis. In this case we found that neutrino self-interactions
decrease the 
electron fraction. In general these self-interaction terms, unlike the
MSW  
effect, tend to transform both neutrinos and antineutrinos. Hence one
expects  
that when the self interaction terms are dominant (e.g when the
neutrino  luminosities are very large) the electron fraction would
reach to the value 
of 0.5. However we showed that, because of the threshold effects in
the  
neutrino interactions, electron fraction exceeds 0.5 even when the 
background neutrinos are numerous. Clearly these conditions are not
favorable  
for the r-process nucleosynthesis. 
We also found that, when background effects are small, with the
parameters we adopted for baryon density and neutrino mixing 
there exists a region 
around 100 km in which electron fraction is smaller than 0.5. Even in
this region alpha effect pulls the value of the electron fraction
closer to 0.5.  
Although the conditions are favorable, the
r-process in this region could not contribute reasonable quantities 
of elements since the baryon density is rather low. 

In contrast, when the initial luminosities of the $\nu_e$'s and
$\bar{\nu}_e$'s are taken to be twice of the other flavors with a much
less pronounced neutrino hierarchy, we found that the asymptotic
electron fraction is less than 0.5 when neutrino self-interactions are
negligible. However, once again the baryon density is low and
r-process in this asymptotic region is likely to produce insignificant
quantities. 
When neutrino self-interactions are dominant the
asymptotic electron fraction shows the same behavior as the first
generic 
case (with a pronounced hierarchy of neutrino energies and
equal luminosities) we considered.  

It is not surprising that the two cases are markedly different for low
neutrino luminosities (late times). If all the initial average 
neutrino energies and luminosities were the same neutrino oscillations
would clearly have no impact. Our analysis highlights the significance
of having a precise knowledge  
of the muon and tau neutrino spectra and luminosities in assessing
core-collapse supernovae as a site of r-process nucleosynthesis. In
both of 
the two generic, but markedly different, cases we studied we found
that neutrino self-interactions are crucial in setting the value of
the electron-fraction to be rather large. We should emphasize that our
treatment of the neutrino self-interactions in Eqs. (\ref{eq:h4}) and
(\ref{eq:h5}) is an approximation using flux-averaged values. Hence 
an exact treatment of the neutrino self-interactions remains to be the
key issue in the proper description of the neutrino transport in
core-collapse supernovae.


\section*{Acknowledgments}

{We thank G. Fuller and G. McLaughlin for useful conversations. ABB 
also thanks the Nuclear Theory Group at Tohoku University for their 
hospitality during the latter part of this work. 
This work was supported in part by the U.S. National Science
Foundation Grant No.\ PHY-0244384 and in part by the University of
Wisconsin Research Committee with funds granted by the Wisconsin
Alumni Research Foundation.}


\end{document}